\documentclass{article}
\usepackage{spconf,amsmath,graphicx}
\usepackage{placeins}
\usepackage{float}
\usepackage{amsfonts}
\usepackage{algorithm2e}
\usepackage[table]{xcolor}
\usepackage{comment}
\usepackage{hyperref}
\newcommand{\nosemic}{\SetEndCharOfAlgoLine{\relax}}

\usepackage{environ}
\usepackage{lipsum} 

\NewEnviron{gather+}[1][1]{%
  \begin{equation}
  \scalebox{#1}{$\begin{gathered}\BODY\end{gathered}$}
  \end{equation}
}

\RestyleAlgo{ruled}

\title{SLICER: Learning universal audio representations using low-resource self-supervised pre-training}
\name{Ashish Seth$^{2\star}$, Sreyan Ghosh$^{1\star}$, S. Umesh$^{2}$, Dinesh Manocha$^{1}$ \thanks{\hspace*{-1mm}$^{\star}$These authors contributed equally to this work}}

\address{
    $^1$University of Maryland, College Park, USA\\
  $^2$Speech Lab, Department of Electrical Engineering, IIT Madras, Chennai, India\\
  }
\begin{document}
%
\maketitle
\begin{abstract}
We present a new Self-Supervised Learning (SSL) approach to pre-train encoders on unlabeled audio data that reduces the need for large amounts of labeled data for audio and speech classification. Our primary aim is to learn audio representations that can generalize across a large variety of speech and non-speech tasks in a low-resource un-labeled audio pre-training setting. Inspired by the recent success of clustering and contrasting learning paradigms for SSL-based speech representation learning, we propose SLICER (\textbf{S}ymmetrical \textbf{L}earning of \textbf{I}nstance and \textbf{C}luster-level \textbf{E}fficient \textbf{R}epresentations) which brings together the best of both clustering and contrasting learning paradigms. We use a symmetric loss between latent representations from student and teacher encoders and simultaneously solve instance and cluster-level contrastive learning tasks. We obtain cluster representations online by just projecting the input spectrogram into an output subspace with dimensions equal to the number of clusters. In addition, we propose a novel mel-spectrogram augmentation procedure \emph{k-mix}, based on mixup \cite{zhang2017mixup}, which does not require labels and aids unsupervised representation learning for audio. Overall, SLICER achieves state-of-the-art results on the LAPE Benchmark \cite{9868132}, significantly outperforming all other prior approaches, sometimes pre-trained on $10\times$ larger unsupervised data  than our setting. Code https: https://github.com/Sreyan88/LAPE.
\end{abstract}
\section{Introduction}
\label{sec:intro}

SSL (self-supervised learning) is increasingly used to obtain good  performance for all modalities corresponding to speech~\cite{baevski2020wav2vec}, vision~\cite{grill2020bootstrap,he2020momentum}, and text ~\cite{devlin2018bert}. In practice, SSL-based speech representation learning has achieved state-of-the-art results in a variety of speech tasks \cite{yang2021superb} like Automatic Speech Recognition (ASR), Phoneme Recognition (PR), etc. However, current SSL-based methods may fail to perform well on tasks that do not involve recognizing phonetic, semantic, or syntactic information in speech~\cite{9868132}, like acoustic scene classification. We hypothesize that this might be due to models learning an implicit language model through solving some form of Masked Acoustic Modeling (MAM) task with individual speech frames (for eg. contrastive learning \cite{baevski2020wav2vec} or clustering \cite{hsu2021hubert}). Learning SSL-based universal audio representations (also known as general-purpose audio representation learning in literature) still remains a relatively nascent area of research and differs from speech representation learning by requiring to learn better global-scale features.
\vspace{0.5mm}

\begin{figure*}[t!]
  \centering
  \includegraphics[width=1.0\textwidth]{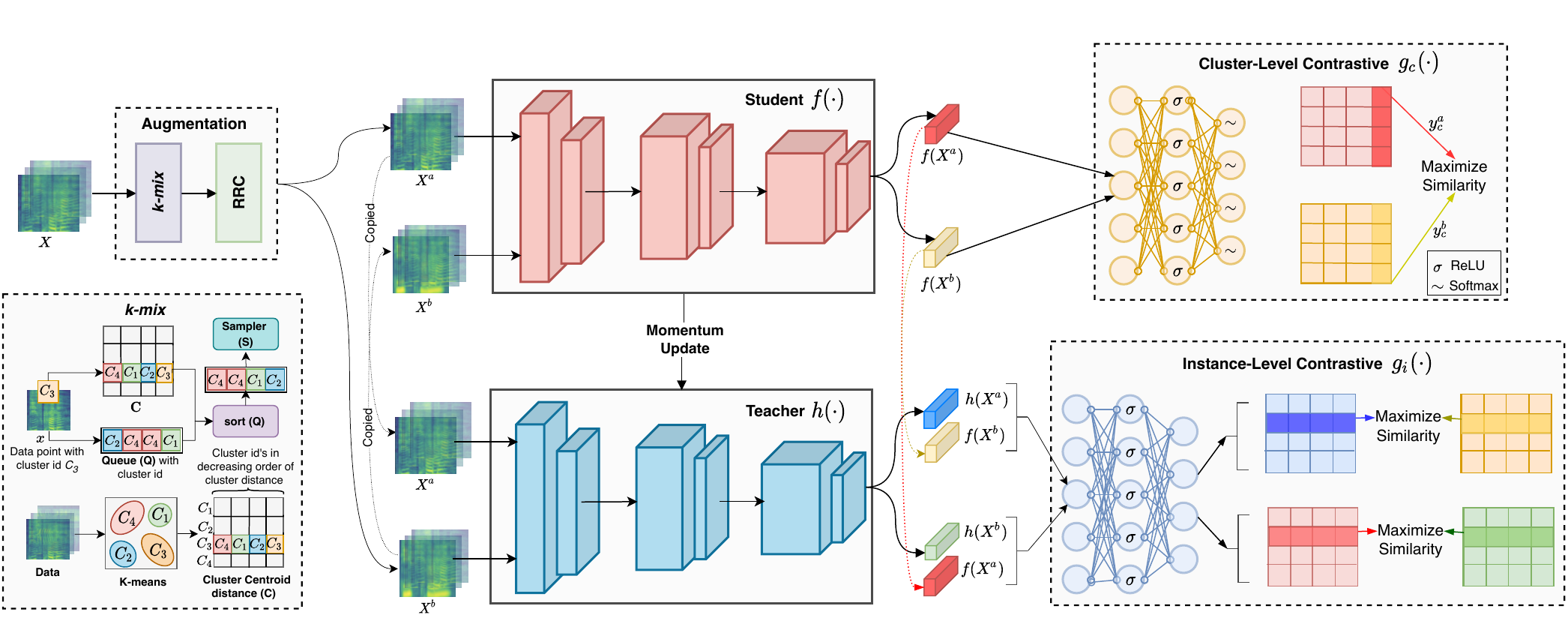}
  \caption{\small Illustration of SSL with \textbf{SLICER} to learn audio representations from un-labeled audio data. \textbf{SLICER} has 2 encoder networks, a student and a momentum-teacher. An unlabeled audio sample is first converted to its log-mel-spectrogram representation before passing it through \textbf{\emph{k-mix}} and RRC augmentations to generate 2 randomly augmented views of the audio sample. Then both these samples are passed through both the encoder networks, and we calculate a cross-contrastive loss between the student output for the first view and the teacher output of the second view and vice-versa. Additionally, the student solves a cluster-level contrastive. 
  Post SSL we evaluate SLICER on 11 speech and non-speech tasks on the \textit{linear evaluation} setup. We also propose \textbf{\emph{k-mix}}, a novel audio augmentation algorithm based on mixup.}
  \label{fig:figure_1}
\end{figure*}

\noindent {\bf Main Contributions:} We present SLICER, a new SSL algorithm for learning general-purpose audio representations from un-labeled audio that simultaneously learns instance-level discriminative features and performs online clustering in a one-stage and end-to-end manner without any extra overhead like an offline clustering step. In general, offline clustering does not scale well with large-scale datasets as they need to perform clustering on the entire dataset at each epoch \cite{ghosh2021deep}. In contrast to prior methods, SLICER learns a deep network that outputs a matrix where rows and columns correspond to the instance and cluster representations, respectively (see Fig. \ref{fig:figure_1}). We achieve this by projecting the input audio log-mel-spectrogram into an output space that is equal to the number of desired clusters centroids. We build this on top of the student and momentum-teacher learning paradigm, where one out of two identical encoders is updated from the momentum of the other \cite{9868132,he2020momentum}. Moreover, for instance-level learning, we use a symmetrical cross-contrastive loss where each encoder calculates a separate loss by sampling negatives for each of the positives from the output of the other. In practice, SLICER outperforms all prior-art on 11 speech and non-speech tasks from the LAPE Benchmark.

\section{Related Work}
\label{sec:format}

Recently, researchers have made considerable progress in devising new and better SSL algorithms for learning general-purpose audio representations. These algorithms solve contrastive \cite{9868132,saeed2021contrastive}, clustering, \cite{ghosh2021deep} or reconstruction tasks \cite{liu2020mockingjay}. To the best of our knowledge, no prior work has successfully solved both contrastive and clustering SSL tasks. Through SLICER, we bring the best of both worlds in an efficient and non-trivial way by solving both contrastive and clustering tasks simultaneously.
Unlike MAM in speech representation learning, most work in SSL for audio representation learning tries to maximize agreement between differently augmented views of the same audio sample \cite{9868132,niizumi2021byol}. This means data augmentations play a critical role in terms of the  effectiveness of the SSL method. Though various methodologies and libraries exist for augmenting raw audio in the time domain \cite{kharitonov2021data}, there is relatively less work on devising better and stronger augmentation schemes for mel-spectrograms. \cite{niizumi2021byol} proposed the use of Random Resized Crop (RRC) and a modified version of mixup, which works in an unsupervised setting without labels. Recently they also proposed the Random Linear Fader, which randomly applies a linear change in volume to the entire audio segment to simulate the approaching or passing away of a sound source or fade-in/out. Our proposed \emph{k-mix} builds on the unsupervised mixup setup proposed by \cite{niizumi2021byol}.

\section{Methodology}
\label{sec:slick}

\subsection{Problem Formulation}
\label{sec:problem_formulation}

Let's suppose $\textbf{X}$ is an unlabeled dataset of size $J$, where $\textbf{X}$ = $\{x_1, \cdots, x_j, \cdots ,x_J\}$ . Here $J$ =  0.25 million, following the exact pre-training setup in the LAPE Benchmark \cite{9868132}. Also let $\textbf{D}^t$ be the task-specific labeled dataset for task $t$ of size $I$. $\textbf{D}^t$ = $\{(x^t_1,y^t_1), \cdots, (x^t_i,y^t_i), \cdots ,(x^t_I,y^t_I)\}$ and $y$ is the corresponding label for audio sample $x$. Our aim here is to use $\textbf{X}$ to pre-train our feature encoder using an SSL algorithm and then fine-tune our model on $\textbf{D}^t$  with supervision, keeping the feature encoder weights frozen.


\subsection{SLICER: Instance-Level Contrastive Learning}
\label{sec:instancecl}

In this section, we describe our instance-level cross-contrastive loss formulation for SLICER. The basic idea of contrastive learning is to identify the true latent representation among of set of distractors or negatives. Following \cite{9868132}, instance-level contrastive learning aims at maximizing the agreement between differently augmented views $x^a_i$ and $x^b_i$ of the same audio sample $x_i \in X$, where $X$ is a batch of size $N$ indexed by $\{0, \cdots, i, \cdots ,N-1\}$. Formally put, we pass each sample $x_i$ in a batch $X$ through a set of augmentations $A$ with some degree of randomness to produce $x^a_i \in X^a$ and $x^b_i \in X^b$. Now each of $X^a$ and $X^b$ is passed through both the student $f(.)$ and teacher encoders $h(.)$ and finally to $g_i(\cdot)$ to obtain embeddings $f(X^a), f(X^b), h(X^a), h(X^b) \in \mathbb{R}^{N \times C}$ and individual instances form rows in the $\mathbb{R}^{N \times C}$ space. We denote these instances as $f(x^a_i), f(x^b_i), h(x^a_i), h(x^b_i) \in \mathbb{R}^{C}$. As mentioned earlier, the SLICER learning paradigm is comprised of two identical student and teacher encoders, where the teacher is updated based on the momentum-based moving average weights of the student. Finally, we calculate the InfoNCE loss as follows:
\vspace{-0.5cm}

\begin{gather+}[0.73]
\label{eqn:info}
\mathcal{L}^{\text {i}}_{\text {InfoNCE}}(f,h)=-\log(
\frac{\exp \left(f\left(x^{a}_{i}\right) \cdot h\left(x^{b}_{i}\right) / \tau\right)}{\exp \left(f\left(x^{a}_{i}\right) \cdot h\left(x^{b}_{i}\right) / \tau\right)+\sum_{i=0}^{K} \exp \left(f\left(x^{a}_{i}\right) \cdot h\left(\tilde{x_{i}}\right) / \tau\right)})
\end{gather+}
While ($f(x^a_i)$,$h(x^b_i)$) make a positive pair, there are $K$ negative pairs between $x^a_i$ and $\tilde{x_{i}}$ where each $\tilde{x_{i}}$ is a randomized augmentation of a different audio sample sampled from $h(x^b)$. Finally we optimize a symmetric cross-contrastive loss $\mathcal{L}^{\text {i}}_{\text {InfoNCE}}$ between the student and the teacher $\mathcal{L}^{\text {i}}_{\text {InfoNCE}}$ = $\mathcal{L}^{\text {i}}_{\text {InfoNCE}}(f,h)$ + $\mathcal{L}^{\text {i}}_{\text {InfoNCE}}(h,f)$ in addition to the cluster-level contrastive loss mentioned in the next section.


\subsection{SLICER: Cluster-Level Contrastive Learning}
\label{sec:majhead}

In this section, we describe our cluster-level contrastive loss formulation for SLICER. As mentioned in the previous section, both our student and teacher networks project the batch of spectrograms into an $N \times C$ dimensional space. Here $C$ represents the number of cluster centroids indexed by $\{0, \cdots, c, \cdots ,C-1\}$ and $N$ represents the number of samples in the batch indexed by $\{0, \cdots, i, \cdots ,N\}$. Formally put, if $g_c(f(X^a)) \in \mathbb{R}^{N \times C}$ is the output of the student for the first augmentation scheme, $g_c(f(X^a))_{n,c}$ can be interpreted as the probability of sample $n$ being assigned to cluster $c$, which can also be interpreted as the ``soft label'' for $n$. Thus, we solve the contrastive learning task on cluster representations in the column space of $g_c(f(X^a))$, where we regard each column as the cluster distribution over each instance. The contrastive learning task tries to solve a task like its instance-level counterpart, where we try to make cluster representations invariant to augmentations applied over the batch of samples. However, differences include the loss being non-symmetrical and only calculated on the output representation space of the student. Thus, considering $y^a_c, y^b_c \in \mathbb{R}^N$ to be the column representations for $g_c(f(X^a)), g_c(f(X^b))$, respectively, we can formulate our contrastive loss as follows:

\begin{gather+}[1.0]
\label{eqn:cont_info}
\mathcal{L}^{\text {c}}=-\log(
\frac{\exp \left(y^{a}_{c} \cdot y^{b}_{c} / \tau\right)}{\exp \left(y^{a}_{c} \cdot y^{b}_{c} / \tau\right)+\sum_{c=0}^{K} \exp \left(y^{a}_{c} \cdot \tilde{y_{c}} / \tau\right)})
\end{gather+}
where $\tilde{y_c}$ is a random negative from the batch. Thus our final loss to be optimized is $\mathcal{L}_{\text {InfoNCE}}$ = $\mathcal{L}^{\text {i}}_{\text {InfoNCE}}$ + $\mathcal{L}^{\text {c}}_{\text {InfoNCE}}$.
\section{Implementation Details}
\label{ssec:implementation}

\begin{table*}[t]

\centering

\caption{Result comparison of various SSL methods with \textbf{SLICER} on the \emph{linear evaluation setup} with frozen encoder. The best results for each task are presented in  \textcolor{blue}{bold}. ``$-$'' signifies that the results were not reported for the task in the original paper.}
\vspace{1mm}
\resizebox{17.83cm}{!}{%
\begin{tabular}{|l c c c c c c c c| c|}
\hline 
\textbf{DT} & \textbf{COLA} & \textbf{BYOL-A} & \textbf{SimCLR} & \textbf{DECAR-V1} & \textbf{DECAR-V2} & \textbf{DeLoRes-S} & \textbf{MoCo} & \textbf{DeLoRes-M} & \textbf{SLICER}\\
\hline
\textit{Speech}& & & & & & & & &\\
\hline

SC-V1 & 71.7 &$-$& 77.3 & 82.3 & 91.6 & 86.1 & $93.6 $ & $ 94.0$ & \cellcolor{blue!9} \textbf{94.8}\\

SC-V2(12) &$-$& 91.0 & 77.2 & 83.0 & 90.6 & 85.4 & $93.2 $ & $ 93.3$ & \cellcolor{blue!9} \textbf{94.2}\\

SC-V2(35) & 62.4 & 92.2 & 66.0 & 73.6 & 87.2 & 80.0 & $89.3 $ & $ 89.7$ & \cellcolor{blue!9} \textbf{90.4}\\

LBS & \cellcolor{blue!9} \textbf{100.0} &$-$& 89.0 & 91.0 & 92.5 & 90.0 & $95.5 $ & $ 95.7$ & 95.7\\

VC & 29.9 & 40.1 & 28.9 & 25.6 & 33.0 & 31.2 & $42.5 $ & $ 45.3$ & \cellcolor{blue!9} \textbf{49.4}\\

IC &$-$&$-$& 59.8 & 63.2 & 65.2 & 60.7 & $65.1 $ & $ 65.2$ & 
 \cellcolor{blue!9} \textbf{66.4}\\

VF & 71.3 & 90.2 & 69.2 & 74.1 & 78.2 & 76.5 & $87.3 $ & $ 88.0$ & \cellcolor{blue!9} \textbf{89.9}\\
\hline
\textit{Non-Speech} & & & & & & & & &\\
\hline

NS & 63.4 & 74.1 & 61.3 & 70.7 & 69.8 & 66.3 & $74.7 $ & $75.0$ & \cellcolor{blue!9} \textbf{76.3}\\

BSD & 77.0 &$-$& 85.2 & 87.7 & 88.5 & 86.7 & $89.0 $ & $ 89.6$ & \cellcolor{blue!9} \textbf{90.0}\\

TUT & $-$ &$-$& 52.4 & 62.5 & 64.6 & 58.6 & $66.7 $ & $ 65.7$ & \cellcolor{blue!9} \textbf{66.8}\\

US8K &$-$& 79.1 & 69.1 & 70.1 & 73.2 & 71.2 & $81.2 $ & $ 82.7$ & \cellcolor{blue!9} \textbf{83.2}\\
\hline
\textbf{Average} & $-$ & $-$ & 66.9 & 71.2 & 75.8 & 72.1 & $79.8 $ & $ 80.4$ & \cellcolor{blue!9} \textbf{81.6}\\
\hline

\end{tabular}%
}
\label{table:without_barlow}

\end{table*}

\begin{algorithm}[t!]
\linespread{0.6}
\caption{SLICER Pre-Training and Evaluation}\label{alg:two}
\CommentSty{// SSL-pretraining}\\
\begin{small}\KwData{dataset $\mathcal{X}$; epoch $\mathcal{E}$; batch size $\mathcal{N}$}\end{small}
\begin{small}
\For{$epoch = 1$ $to$ $\mathcal{E}$}{
  \nosemic sample a mini batch $\{x_i\}_{i=1}^N$ from $\mathcal{X}$\;
  \nosemic sample two set of augmentations $\mathcal{X}^a$ and $\mathcal{X}^b$\;
  \nosemic compute \textbf{symmetric instance level contrastive loss} $\mathcal{L}^{\text {i}}_{\text {InfoNCE}}$ = $\mathcal{L}^{\text {i}}_{\text {InfoNCE}}(f,h)$ + $\mathcal{L}^{\text {i}}_{\text {InfoNCE}}(h,f)$ (use Eq: \ref{eqn:info})\;
  \nosemic compute \textbf{cluster level contrastive loss} $\mathcal{L}^{\text {c}}_{\text {InfoNCE}}$ (use Eq: \ref{eqn:cont_info})\;
  \nosemic compute $\mathcal{L}_{\text {InfoNCE}}$ = $\mathcal{L}^{\text {i}}_{\text {InfoNCE}}$ + $\mathcal{L}^{\text {c}}_{\text {InfoNCE}}$\;
  \nosemic update $f$, $g_i$ and $g_c$ through gradient descent to minimize $\mathcal{L}_{\text {InfoNCE}}$\;
  \nosemic update $h$ using momentum update from $f$.\; 
}
\CommentSty{// Linear Evaluation ($f$ is frozen)}\\
\KwData{task specific dataset $\mathcal{D}^t$; epoch $\mathcal{E}$; batch size $\mathcal{N}$}
\For{$epoch = 1$ $to$ $\mathcal{E}$}{
\nosemic sample a mini batch $X^t,Y^t$ = $\{(x^t_i,y^t_i)\}_{i=1}^N$ in $\mathcal{D}^t$\;
\nosemic compute latent representation $L^t$ = $f(X^t)$\;
\nosemic compute cross entropy $\mathcal{L}_{ce}(Y^t,\hat{Y}^t)$ where $\hat{Y}^t$ is $\hat{Y}^t$ = $softmax(l(L^t))$ and $l$ is $Linear(\cdot)$ transformation \;
\nosemic update $l$ through gradient descent to min $\mathcal{L}_{ce}$ \;
}
\end{small}

\end{algorithm}

\subsection{Data Augmentation}
\label{sssec:dataaug}

Contrastive learning has historically been known to depend on the quality of augmentations, and for our setup, we use Random Resized Crop (RRC) and \emph{k-mix} as our set of augmentation functions $A$. In the next subsection, we describe in detail our proposed \emph{k-mix} algorithm.
\vspace{1.0pt}

{\noindent \textbf{\emph{k-mix:}}} It's our novel augmentation algorithm for augmenting audio samples. \emph{k-mix} is inspired by the original mixup \cite{zhang2017mixup} proposed for images and its modification for log-mel-spectrograms augmentation in the absence of labels proposed by \cite{niizumi2021byol}. With a log-mel-spectrogram as an input, mixup mix past randomly selected input audio in a small ratio \cite{niizumi2021byol}. As a result, added audio becomes a part of the background sound in the mixed audio. Thus, the mixup augmentation operation is defined as $\tilde{x}_{i}=\log \left((1-\lambda) \exp \left(x_{i}\right)+\lambda \exp \left(x_{k}\right)\right)$ , where $x_k$ is a mixing counterpart and the mixing ratio $\lambda$ is sampled from uniform distribution $U(0.0, \alpha)$ instead of from a beta distribution in the original mixup. In addition, $\alpha$ is a mixing ratio hyper-parameter that controls the degree of contrast between the resulting two mixed outputs. $x_k$ is randomly chosen from a memory bank, which is a FIFO queue storing past inputs; this is where \emph{k-mix} tries to sample audio from the memory bank, which would result in a stronger augmentation. 

We acknowledge that the unlabeled audio dataset used for SSL might not be from largely diverse sources, and thus we hypothesize that randomly sampling audio from a FIFO queue for background noise might result in weak augmentations if the audio is from the same hidden class or source. Thus, kmix tries to sample audios from the queue, and these audio samples are further apart in the Euclidean space, which can be identified in an unsupervised manner using clustering. Formally put, we first train a simple k-means clustering algorithm on about 10\% of our unlabeled AudioSet dataset to obtain $k$ cluster centroids, where $k_i \in R^{128}$ (spectrogram with frequency dimension 128 mean-pooled across time axis). The $k$ cluster centroids are then used to calculate a $k \times k$ matrix $\mathbf{C}$ where $c_{i,j} \in C$ represents the euclidean distance between the $i^{th}$ and $j^{th}$ cluster centroids. Thus, as shown in Fig. \ref{fig:figure_1}, when a new audio sample $x$ is to be augmented, we first find the closest cluster centroid among $k$, to $x$ and all other samples in the FIFO queue. Then using $\mathbf{C}$, we sort the FIFO queue $\mathbf{Q}$ in descending order of distance to the sample $x$ with respect to the cluster centroids that $x$ and samples in $\mathbf{Q}$ belong to. Finally, our sampler $\mathbf{S}$ randomly samples an audio segment from the first $r$ samples of the sorted FIFO queue to augment our incoming audio sample.

\subsection{Experimental Setup}
\label{sec:page}

The primary focus of our work is to devise better SSL algorithms for audio representation learning, and thus we do not focus on devising better architectures or measuring the effect of SSL approaches across different architectures. Thus following much of prior art \cite{9868132,niizumi2021byol}, for both our encoders (student and teacher), we borrow the simple yet powerful architecture proposed by \cite{niizumi2021byol}. For more details, we refer our readers to \cite{9868132,niizumi2021byol}. For a fair comparison, except COLA, which uses EfficientNet, in Table \ref{table:without_barlow} (BYOL-A also uses the same encoder as \cite{niizumi2021byol}), we reproduce results for SSL methodologies in literature with our encoder if they were originally presented with a different architecture. For \emph{k-mix} we find optimal values of $k$ = 128, $r$ = 128, and length of $\mathbf{Q}$ = 2048. For SSL pre-training, we use an embedding dimension $C$ = 256, a learning rate of $3e^{-4}$, a batch size of 1024, and train it for 100 epochs.


\subsection{Datasets}
\label{sec:datasets}

In our experiments, we use the exact same upstream and downstream training setups proposed by LAPE \cite{9868132}. For SSL-based pre-training, we use a balanced subset of 10\% of the complete AudioSet (0.2 million) and the FSD50K \cite{fonseca2021fsd50k}. For downstream tasks (DT), we evaluate our learned representations on LibriSpeech (LBS) \cite{7178964} and VoxCeleb (VC) \cite{Nagrani_2017} for speaker identification, Speech Commands (SC) v1 and v2 \cite{warden2018speech} for keyword spotting, VoxForge (VF) \cite{Voxforge.org} for language identification,  IEMOCAP (IC) \cite{busso2008iemocap} for speech emotion recognition, NSynth \cite{engel2017neural} for TUT Urban \cite{mesaros2018multidevice} and US8K \cite{10.1145/2647868.2655045} for acoustic event classification and finally Bird Song Detection (BSD) \cite{stowell2019automatic}.

\section{Results and Result Analysis}
\label{sec:results}

As we see in Table \ref{table:without_barlow}, SLICER outperforms all other approaches in literature by a significant margin. Results of COLA and BYOL-A were borrowed from their original papers. SimCLR was proposed as the pre-training approach in \cite{wang2022towards}. We attribute the gap in results from the original paper to the powerful encoder proposed in the paper. However, as stated earlier, measuring the effect of change in encoders is beyond the scope of this paper. MoCo inspired from \cite{he2020momentum}, can be viewed as SLICER without symmetric cross-contrastive instance-level learning and cluster-level contrastive learning. Table \ref{table:ablations} shows ablations on various novel components in SLICER. Starting from MoCo proposed in \cite{9868132}, we get 0.4\% average boost by first introducing symmetricity in a cross-contrastive setting, followed 1.2\% on adding cluster-level contrastive and finally another 0.4\% with \emph{k-mix}.

\begin{table}[h]
\caption{Ablation on various components in SLICER.}
\vspace{1mm}
\label{table:ablations}
\centering
\begin{tabular}{|l c|}
\hline 
\textbf{Method} & \textbf{Avg. Accuracy}\\
\hline
MoCo & 79.8 \\
+ symmetric cross-contr. & 80.2 \\
+ cluster contr. (SLICER) & 81.2 \\
+ \emph{k-mix} & \cellcolor{blue!9} \textbf{81.6} \\
\hline

\end{tabular}
\end{table}

\vspace{-1mm}
\section{Conclusion}
\label{sec:conclusion}

In this paper, we propose SLICER, a novel methodology to learn general-purpose audio representations from un-labeled data on low-resource data regimes. SLICER significantly outperforms all other approaches in literature, sometimes even systems trained on $10\times$ more unlabeled data than our setup. We also propose \emph{k-mix} a new log-mel-spectrogram augmentation algorithm that improves over the widely used \cite{zhang2017mixup}.

\vfill\pagebreak

\bibliographystyle{IEEEbib}
\bibliography{strings,refs}

\end{document}